\documentclass[
aps,%
11pt,%
final,%
notitlepage,%
oneside,%
twocolumn,%
nobibnotes,%
nofootinbib,%
superscriptaddress,%
noshowpacs,%
centertags]%
{revtex4}
\begin{document}

%\keywords{galaxies: spiral --- catalogs ---surveys}

\title{Visual Survey of 18\,020 Objects from the 2MFGC Catalog}

\author{\firstname{S.~N.}~\surname{Mitronova}}
\email{mit@sao.ru} 
%\firstaffiliation{\saoname}
\author{\firstname{G.~G.}~\surname{Korotkova}}
\email{ggk@sao.ru}
\affiliation{Special  Astrophysical  Observatory,  Russian  Academy  of  Sciences, Russia}
\received{October 2, 2014}  \revised{November 5, 2014}
\onecolumngrid
{\scriptsize ISSN 1990-3413, Astrophysical Bulletin, 2015, Vol. 70, No. 1, pp. 24--32. c Pleiades Publishing, Ltd., 2015.
Original Russian Text S.N.Mitronova, G.G.Korotkova, 2015, published in Astrofizicheskii Byulleten,
 2015, Vol. 70, No. 1, pp. 24--33.}

\begin{abstract}
We conducted a continuous survey of infrared and visual images of
18\,020 2MFGC galaxies which were selected on an automatic basis
from 1.64~mln extended objects of the 2MASS~XSC catalog based on
the ratio of the infrared axes $a/b\geq3$. This work aims to
exclude ``false'' objects from the list of flat galaxies. Having
observed more than 80 thousand images in different filters, we
were able to detect 1512 such objects (8.4\% of the total number).
We found 23~galaxies duplicated in 2MASS, which have two 2MFGC
numbers correspondingly, and three flat galaxies which are not
included in other catalogs and are located close to three
``false'' galaxies. Galaxies with magnitudes fainter than $K_s =
13^{\rm m}$ compose the main part of the excluded objects. They
show small angular sizes, low surface brightnesses and
concentration ratios. The results of the work in the form of the
2MFGC table with notes are given in the astronomical databases
VizieR,\footnote{\tt http://cdsarc.u-strasbg.fr/viz-bin/qcat?J/other/\\/AstBu/70.24}
NED, HyperLeda.
\end{abstract}
\maketitle

\section{INTRODUCTION}

Nowadays, the researchers involved in the studies of different
cosmological parameters of the Universe, large-scale peculiarities
of the galaxy distribution, their collective motions, and also
structural features of disk galaxies direct their attention to the
spiral edge-on galaxies with thin disks.
%Spiral edge-on galaxies with thin disks attract researchers'
%attention when studying different cosmological parameters of the
%Universe, large-scale peculiarities of the galaxy distribution,
%their collective motions, and also structural features of disk
%galaxies.
In 1993 the catalog of 4455~flat spiral edge-on galaxies,
FGC~\cite{kar1993:Mitronova_n}, was published; it was compiled
based on a continuous review of blue and red images from the
POSS-I and ESO/SERC sky surveys. It consists of two parts: FGC
(Flat Galaxy Catalogue) itself, covering the region \mbox {$\delta
>-20\degr$}, and its southern extension FGCE (Flat Galaxy Catalogue
Extension), $\delta < -20\degr$. The catalog includes objects with
the maximum angular diameter $a_{\rm lim}=0\farcm6$ and the axial
ratio $a/b\geq7$, where $a$ and $b$ are the major and minor axes
respectively. Eventually, in the Digital Sky Survey (DSS), the
coordinates of all the objects were measured again with an
accuracy of up to $3\arcsec$; the diameters determined with the
films from the ESO/SERC survey were adjusted to the
\mbox{POSS-I}~\cite{kud1997a:Mitronova_n} diameter system; the
total visible $B_t$ magnitudes were derived, which correspond to
the $B_t$ values from RC3~\cite{kud1997b:Mitronova_n} with an
accuracy of up to $0\fm25$. As a result, the improved and extended
edition of the RFGC catalog (Revised Flat Galaxy
Catalogue)~\cite{kar1999:Mitronova_n} was released; it comprised
4236 flat galaxies with the extreme axial ratio $(a/b)_{\rm
lim}=7$. The galaxies with reduced diameters $a<0\farcm6$ were
excluded from the RFGC list. At present, the RFGC catalog is
frequently used for studying the characteristics of the star
formation rate, large-scale flows of galaxies, structural features
of disk systems,
etc.~\cite{kar1989:Mitronova_n,kar2000:Mitronova_n,kar2002:Mitronova_n,biz2002:Mitronova_n,biz2009:Mitronova_n,zasov2002:Mitronova_n}.

Current deep sky surveys provide new possibilities for the
detection of objects of such type. Since 2006, using the
observational data from a deep sky survey in the visible ($u, g,
r, i, z$) range---Sloan Digital Sky Survey
(SDSS)~\cite{york2000:Mitronova_n}---a catalog of \mbox{edge-on}
disk galaxies is being compiled on an automatic
basis~\mbox{\cite{Kaut2006a:Mitronova_n,Kaut2006b:Mitronova_n,Kaut2009:Mitronova_n,biz2014:Mitronova_n}}.
By now, the SDSS survey covers a quarter of the sky, and 5747
edge-on galaxies have been found in this area.

As is well known, global cosmological investigations require a
homogeneous sample of galaxies all over the sky with sufficient
spatial depth and accuracy of both the coordinates and the
measured values. The best all-sky surveys in terms of the listed
parameters are those conducted in the same manner and with similar
instruments and devices. By the end of the year 2000, such surveys
were the Two Micron All-Sky Survey
(2MASS)~\cite{cur1998:Mitronova_n} and the Extended Sources
Catalog (XSC)~\cite{skr1997:Mitronova_n} compiled on its basis.
%By the end of the year 2000, the Two Micron All-Sky Survey
%(2MASS)~\cite{cur1998:Mitronova_n} and also the Extended Sources
%Catalog (XSC)~\cite{skr1997:Mitronova_n} compiled on its basis
%were such surveys.
One of the most significant advantages of this
catalog, and also the main reason why the 2MASS survey was
conceived, is the low absorption of our Galaxy in the infrared
(IR) range compared to the visible one. Thus, in the IR-range the
Galaxy becomes more transparent for the search for remote objects
in the direction of its stellar disk~\cite{kar1999:Mitronova_n},
which prompted us to compile a catalog of flat galaxies selected
from 2MASS.

The objects for the catalog of disk-like galaxies---the
2MASS-selected Flat Galaxy Catalog\linebreak
(2MFGC)~\cite{mit2004:Mitronova_n}, comprising 18\,020 objects all
over the sky---were automatically selected from 1.64~mln extended
objects of 2MASS~XSC. Among the objects with \mbox {$K_s<14^{\rm
m}$} and angular diameters larger than~7$\arcsec$, we selected the
objects with axial ratios $b/a\leq0.34$ or $a/b\geq3$, which
corresponds to the visible axial ratio $a/b\geq6$. The axial
ratios both on the combined \mbox {$J+H+K_s$} image ($sba$) and in
each filter were taken into consideration.
%Moreover, we took into consideration the axial ratios both on the
%combined \mbox {$J+H+K_s$} image ($sba$) and in each filter.
Our choice of this criterium was driven by the comparison of the
infrared and optical characteristics of the RFGC
galaxies~\cite{kar2002:Mitronova_n}.
%We chose this criterium based on the comparison of the infrared
%and optical characteristics of the galaxies from the
%RFGC~\cite{kar2002:Mitronova_n}.

The main goal of compiling the flattened galaxy catalog is to
obtain the deepest, morphologically homogeneous sample of spiral
field galaxies across the sky. Unlike other available optical
catalogs, e.g., the RFGC~\cite{kar1999:Mitronova_n}, the 2MFGC
catalog seems to be more relevant for the studies of cosmic
streams on a scale of $z\la0.1$. For example, it was shown
in~\cite{mit2004:Mitronova_n} that the dipole moment of the
distribution of bright (\mbox {$K<11^{\rm m}$}) 2MFGC objects
(\mbox {$l=227\degr$}, $b=41\degr$) is within the statistical
error ($\pm15\degr$) in the direction of the IRAS dipole and the
optical RFGC dipole.

It is known that the high brightness of the night sky in the
near-IR region and the short exposures (about 8~s/object) of
2MASS~\cite{jar2000:Mitronova_n} make the selection of extended
sources difficult. As a result, the periphery of spiral galaxy
disks is not usually seen in the isophotes fainter than $K_s$ =
$20^{\rm m}/\sq\arcsec$. This, in turn, results in the fact that,
on the one hand, bright galaxies with low surface brightness and
late-type morphology can be omitted from the XSC catalog. It is
shown in~\cite{kar2002:Mitronova_n} that from 4236 edge-on \mbox
{RFGC galaxies}, only 2996 (71\%) were detected in 2MASS, and the
data on them are available in the XSC catalog. About 18\% of these
2996 RFGC galaxies are not included in the 2MFGC catalog, as their
axial ratios ($b/a\geq0.34$) are beyond the selection criterium
limits of the 2MFGC. On the other hand, for the same reason
``false'' objects may be included in the XSC: the result of
overlapping when performing the photometry of a pair or a chain of
galaxies or stars, a galaxy and a projected star (stars), and also
galaxies with an elongated red bar or a bulge, the spiral
structure of which is detectable only in the visible range. To
reduce the influence of such errors, in the course of compiling
the 2MFGC catalog we reviewed several thousand images of galaxies
on the $J, H, K_s$ frames from the 2MASS and DSS1 catalogs.
However, in the course of time the necessity of total revision of
the images of 2MFGC objects became obvious. Generally, this work
aims to exclude ``false'' objects in order to improve the accuracy
of future investigations.

\section{VIEWING TECHNIQUE FOR INFRARED AND VISIBLE IMAGES OF 2MFGC~OBJECTS}

A continuous review of available images of\linebreak 2MFGC objects
was conducted with the $J, H, K_s$\linebreak \mbox{2MASS frames}
and their sum in the NED\footnote{\tt http://ned.ipac.caltech.edu}
and DSS2-red database systems; we used the\linebreak
\mbox{DSS2-blue}, DSS2-infrared and DSS1\footnote{\tt
http://archive.eso.org/dss/dss/} for refining, and it was possible
to use the combined images from the Sloan Digital Sky Survey (SDSS
III) DR9\footnote{\tt
http://skyserver.sdss3.org/dr9/en/tools/chart/\\/navi.asp} for
about a quarter of the objects. Eventually, we provided the 2MFGC
table with notes and placed it in the astronomical electronic
databases VizieR,\!\label{maintable:Mitronova_n}
NED, HyperLeda.

Figures~1 and~2 show examples of several infrared and visible
galactic images left in the catalog (Fig.~1) and the objects which
were, for one reason or another, wrongly classified as flat
galaxies (Fig.~2). The combined \mbox {$J + H + K_s$} 2MASS images
are given on the left, the DSS2 or SDSS images of the same objects
are shown on the right.

\begin{figure*}
%\vspace{1mm}
%\begin{minipage}{1.0\linewidth}
\setcaptionmargin{2mm} \onelinecaptionsfalse \captionstyle{normal}
\includegraphics[scale=0.4]{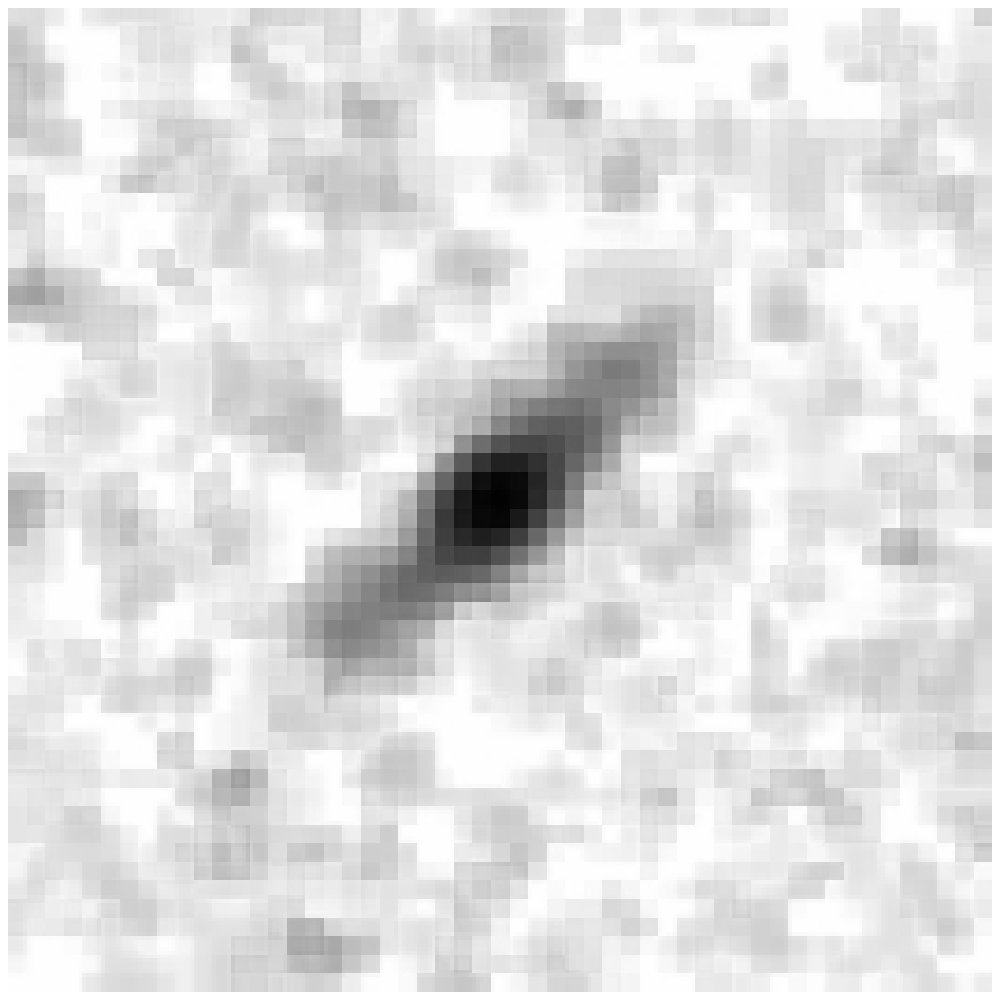} \hspace{0.2cm}
\includegraphics[scale=0.4]{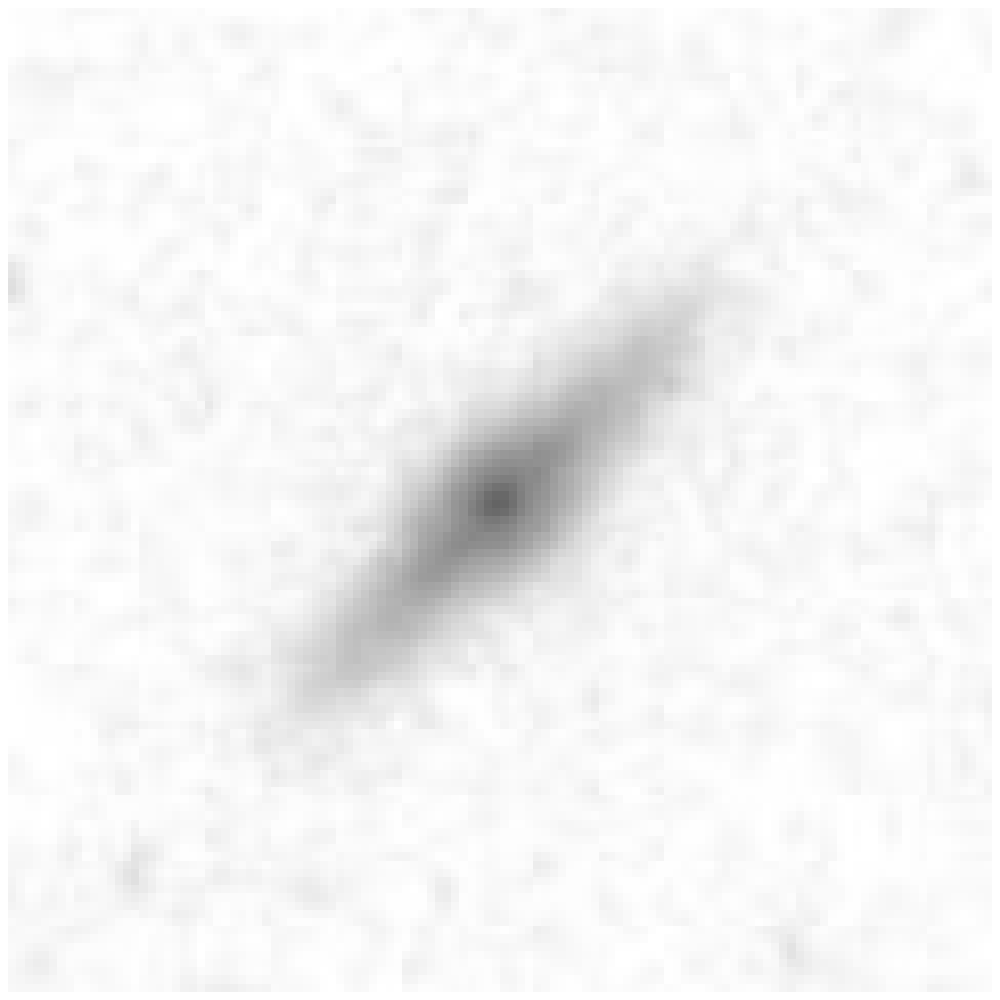}\\
{2MFGC\,813}\vspace{0.5mm}
\\
\includegraphics[scale=0.5]{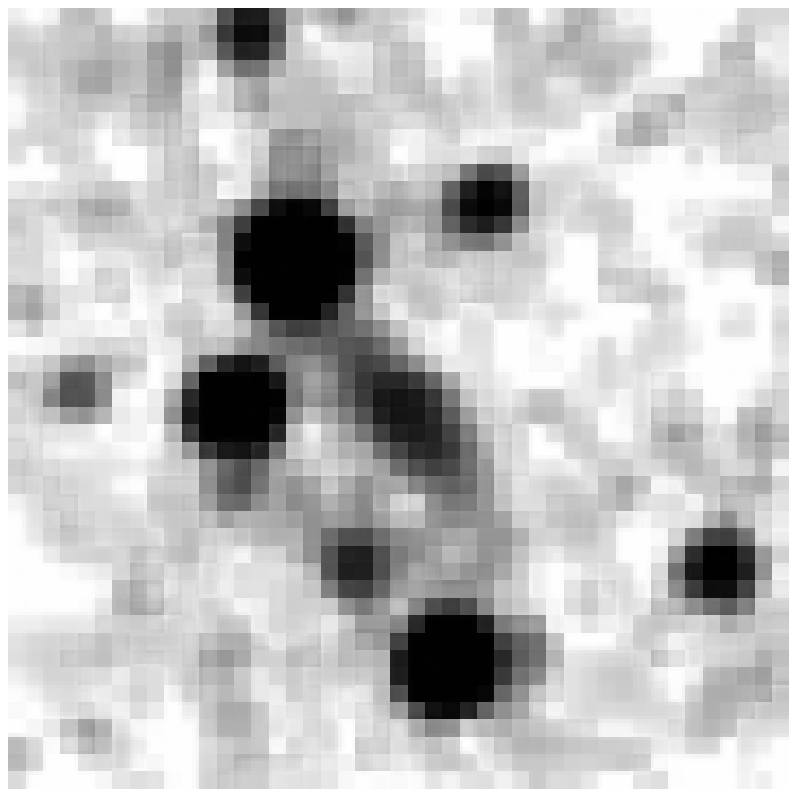}\hspace{0.2cm}
\includegraphics[scale=0.4]{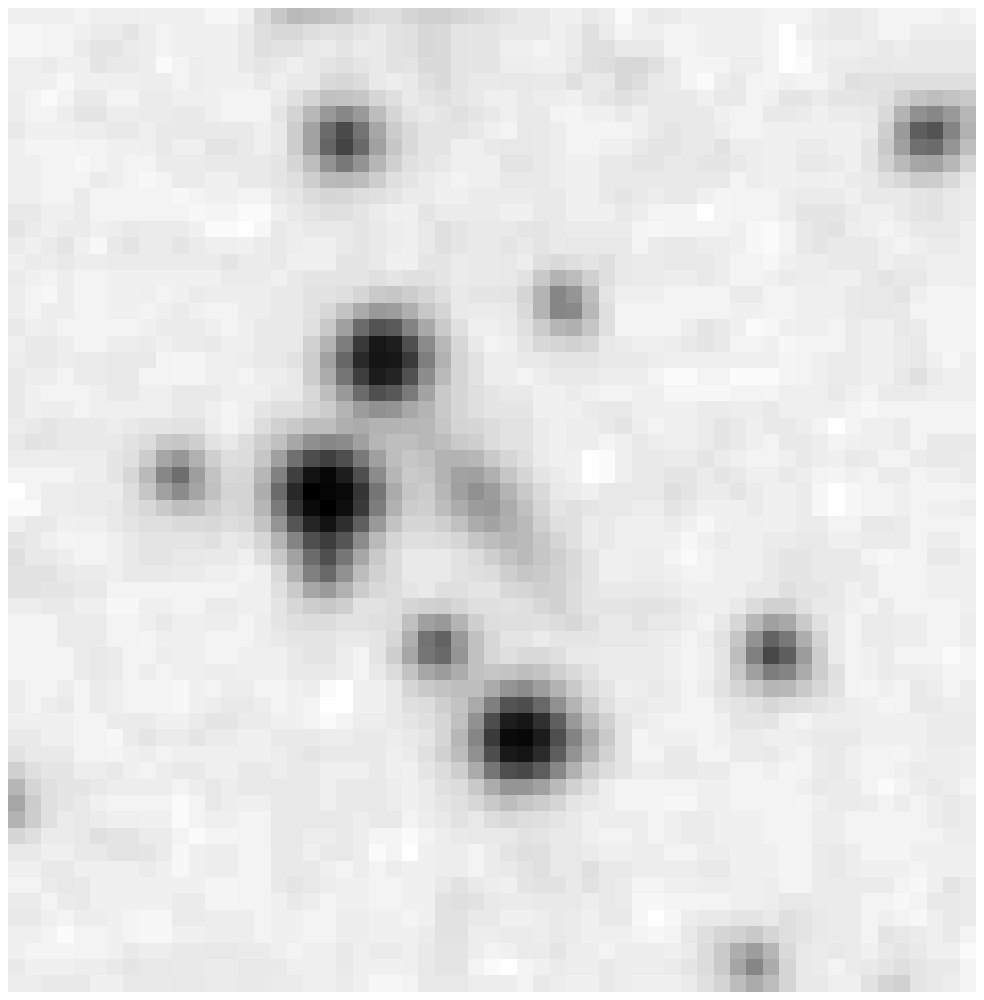}\\
{2MFGC\,895 ($b = -5\fdg5$)}\vspace{0.5mm}
\\
\includegraphics[scale=0.4]{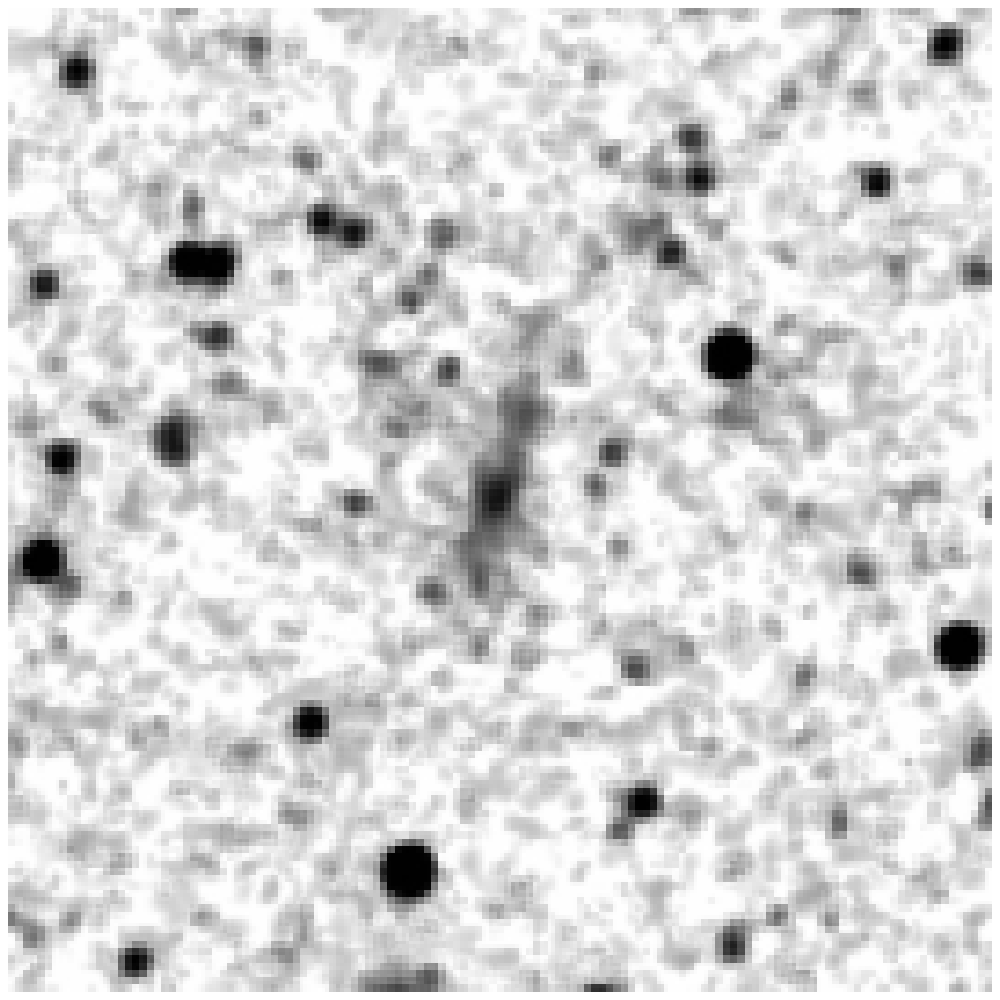}\hspace{0.2cm}
\includegraphics[scale=0.4]{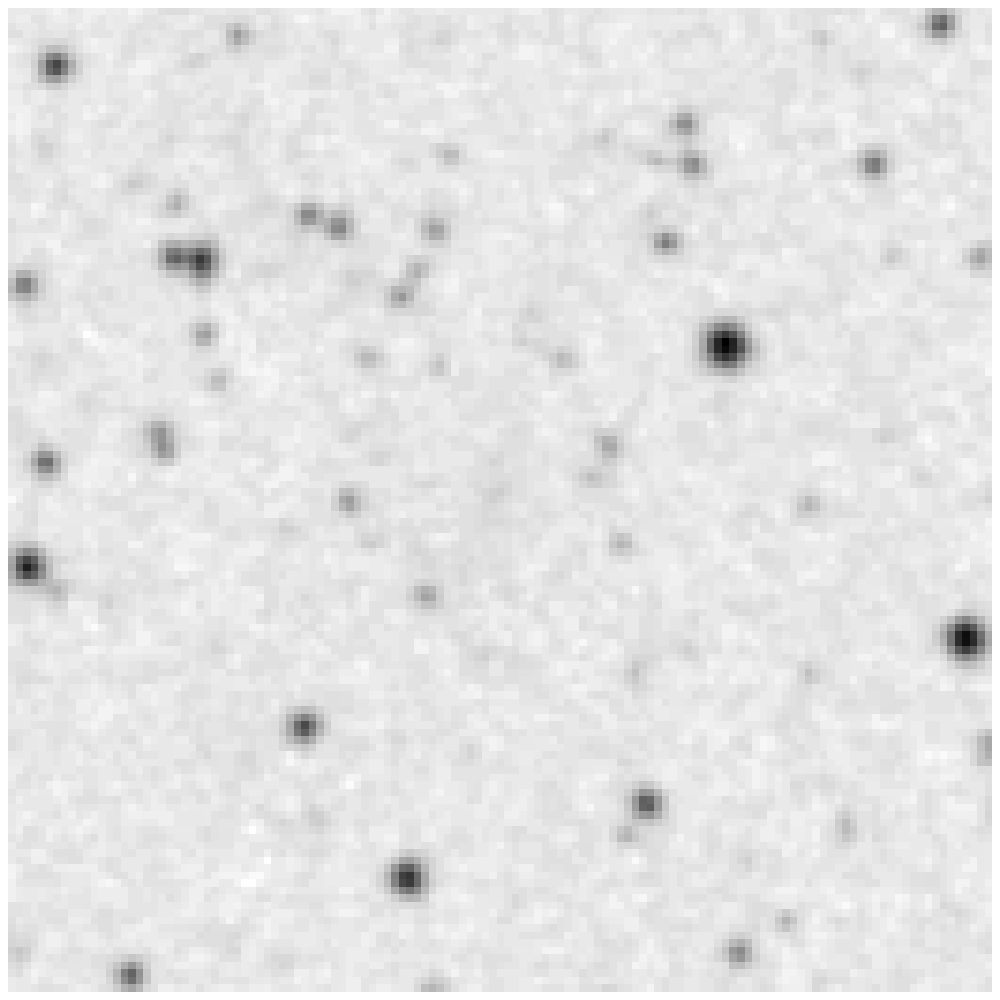}\\
{2MFGC\,1119 ($b = 0\fdg6$)}\vspace{0.5mm}
\\
\includegraphics[scale=0.4]{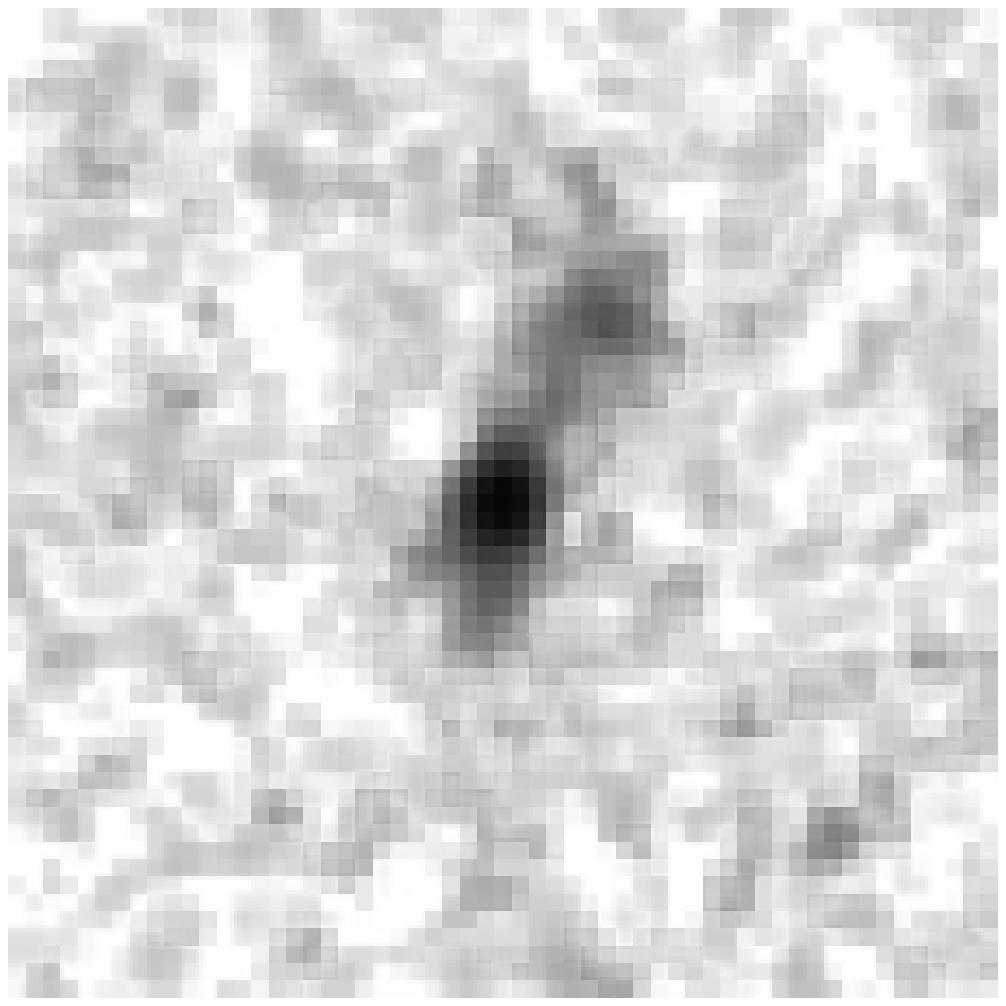}\hspace{0.2cm}
\includegraphics[scale=0.4]{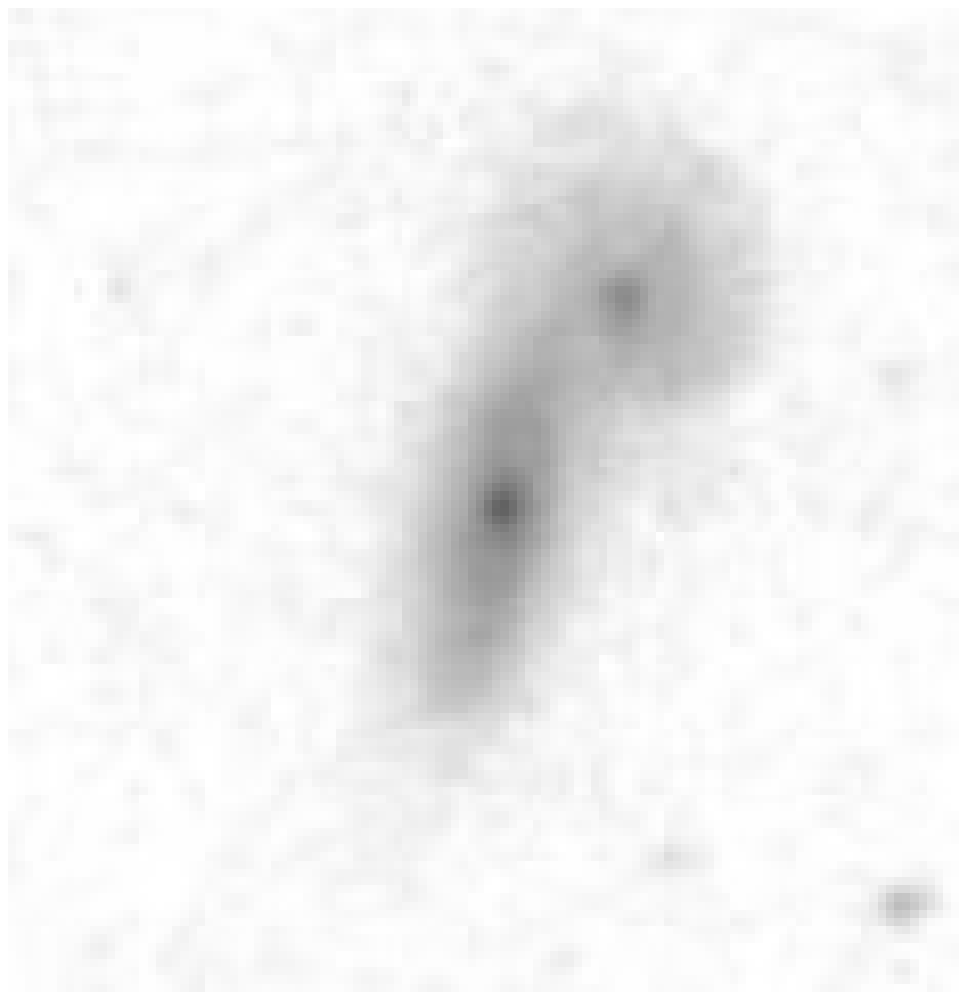}\\
{2MFGC\,9497}
 \caption{Examples of confirmed galaxies from 2MFGC.
The combined $J + H + K_s$\,\, 2MASS images (on the left) and the
DSS2 or SDSS images (on the right).}
\end{figure*}
%\end{minipage}
%\clearpage
\hfill
%\begin{minipage}{1.0\linewidth}
\begin{figure*}
 %\vspace{2mm}
\setcaptionmargin{5mm} \onelinecaptionsfalse \captionstyle{normal}
\includegraphics[scale=0.4]{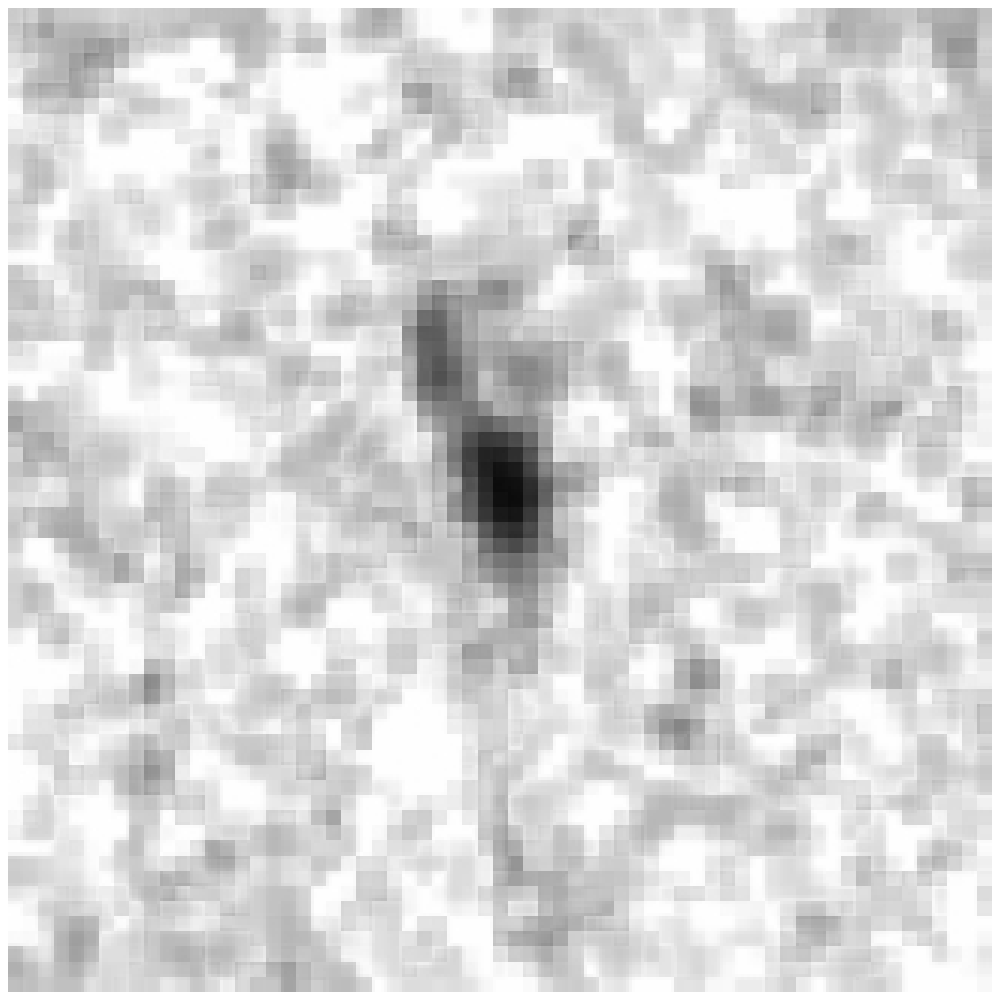}\hspace{0.2cm}
\includegraphics[scale=0.4]{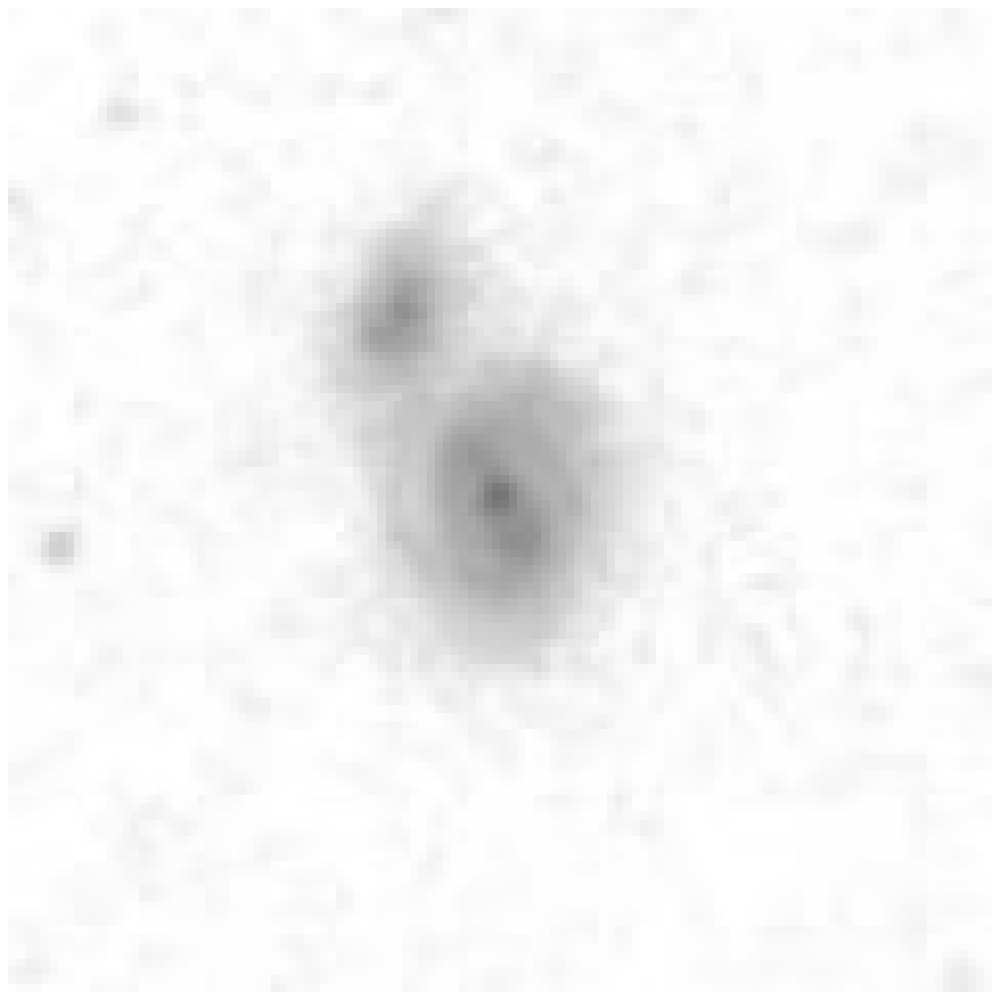}\\
{2MFGC\,151}\vspace{0.5mm}
\\
\includegraphics[scale=0.4]{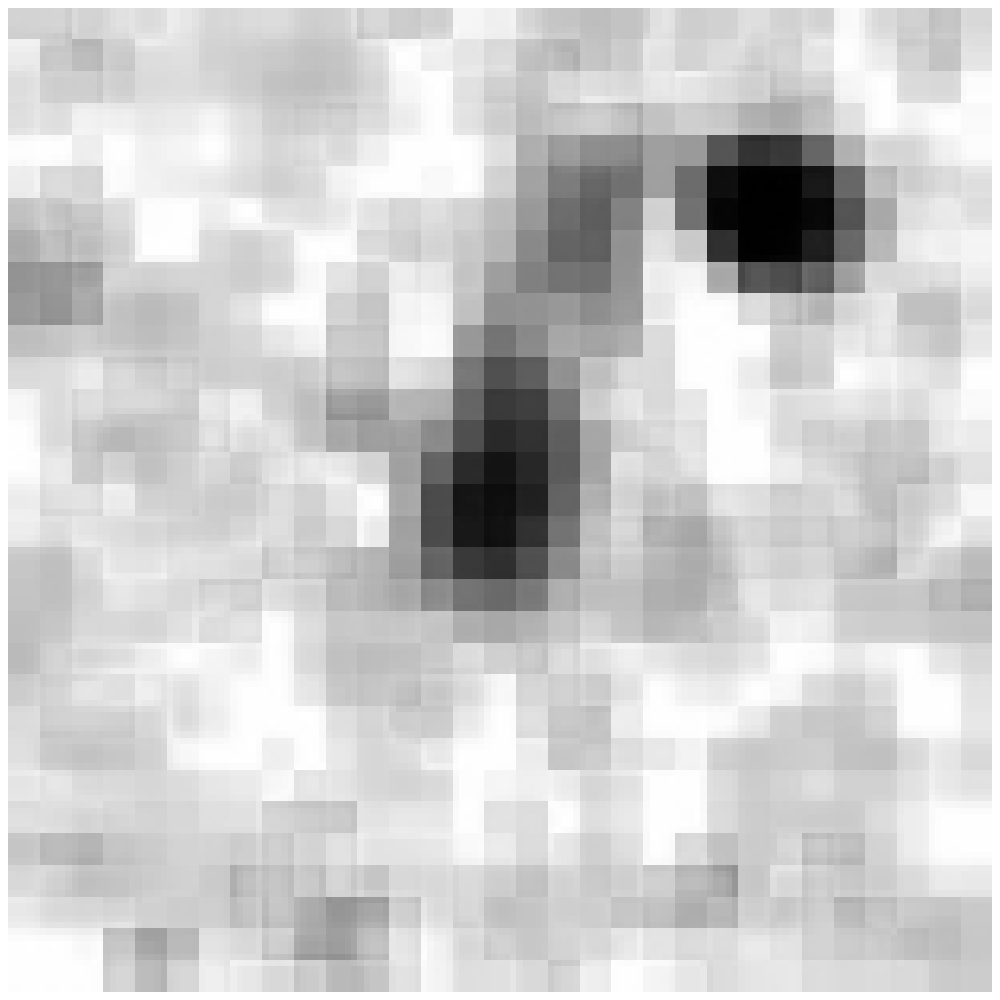}\hspace{0.2cm}
\includegraphics[scale=0.4]{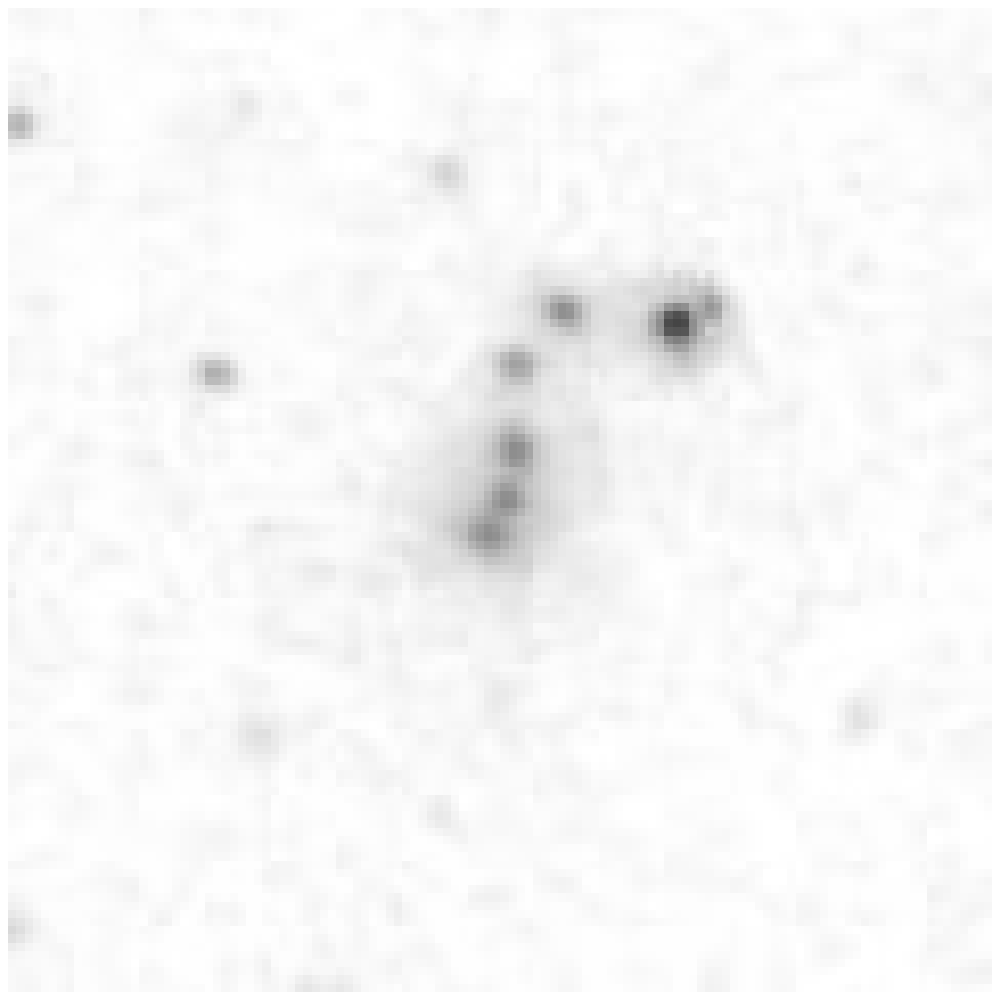}\\
{2MFGC\,673}\vspace{3.0mm}
\\
\includegraphics[scale=0.4]{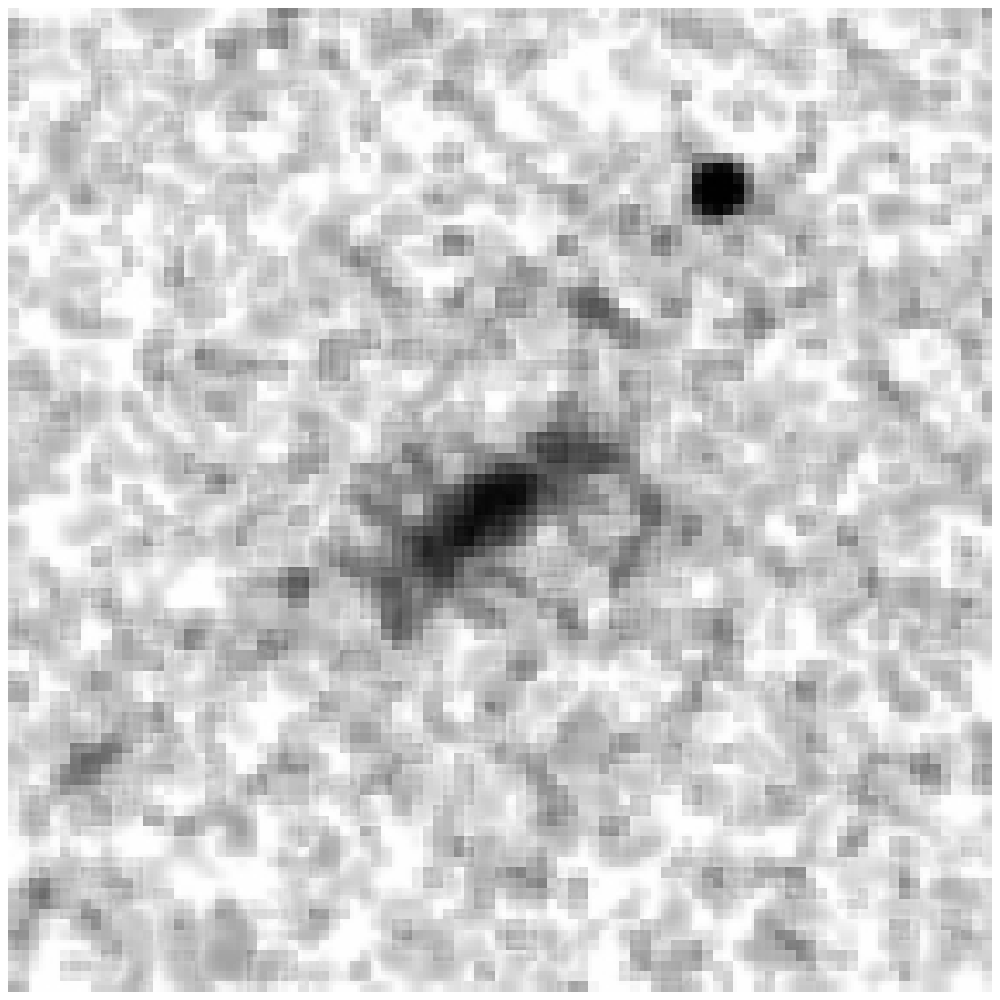}\hspace{0.2cm}
\includegraphics[scale=0.4]{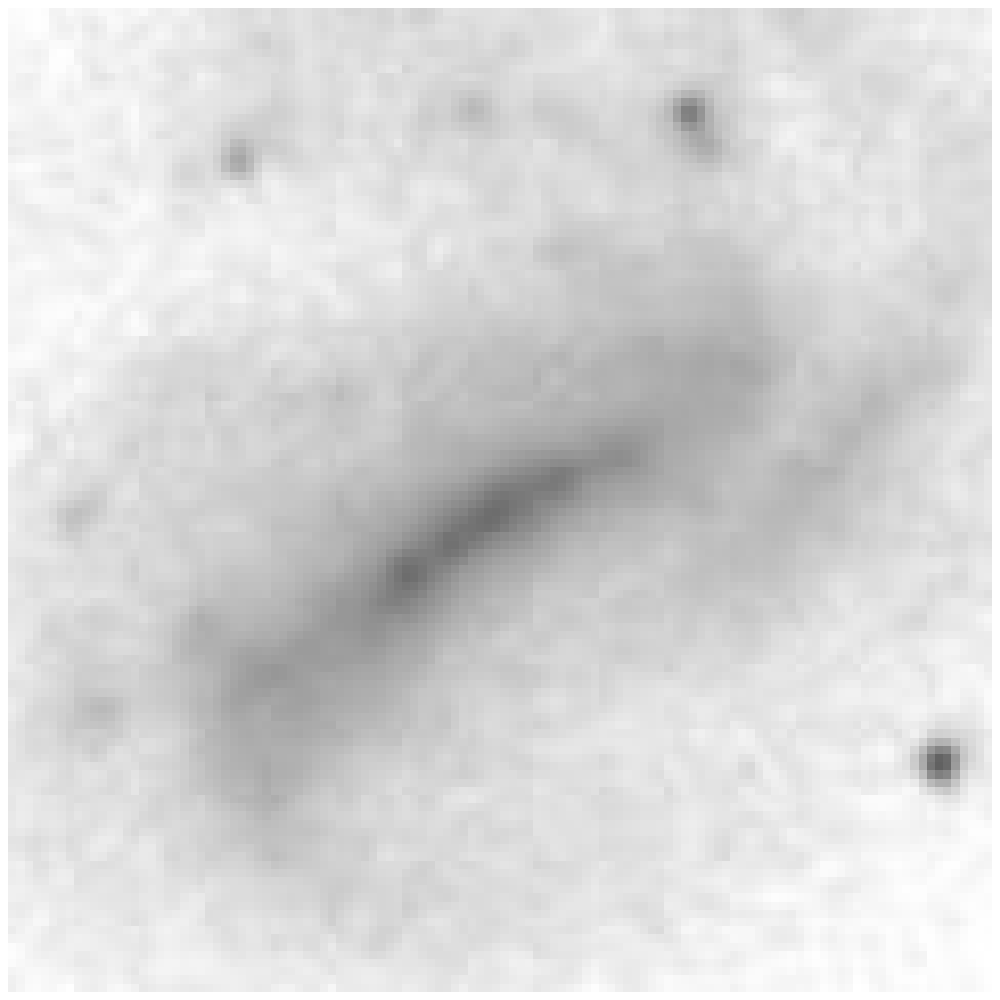}\\
{2MFGC\,5959}\vspace{3.5mm}
\\
\includegraphics[scale=0.4]{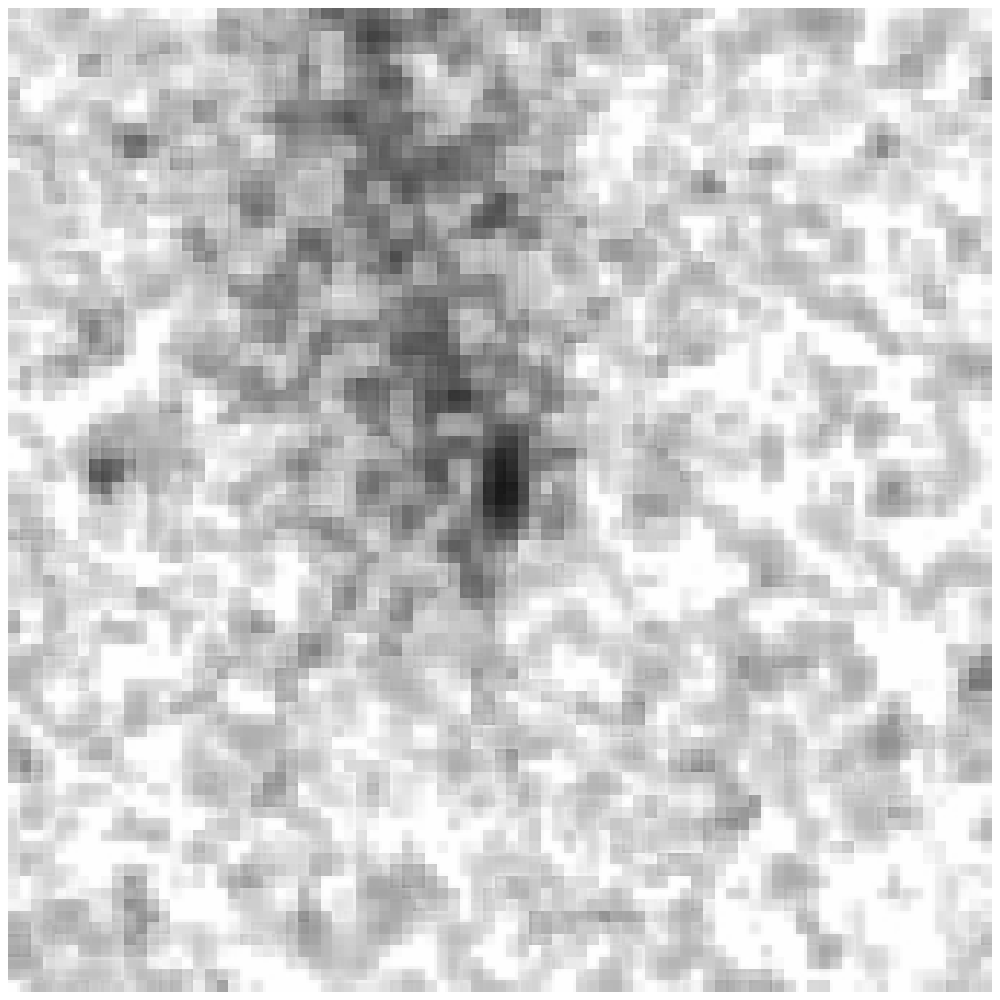}%\hspace{0.2cm}
\includegraphics[scale=0.4]{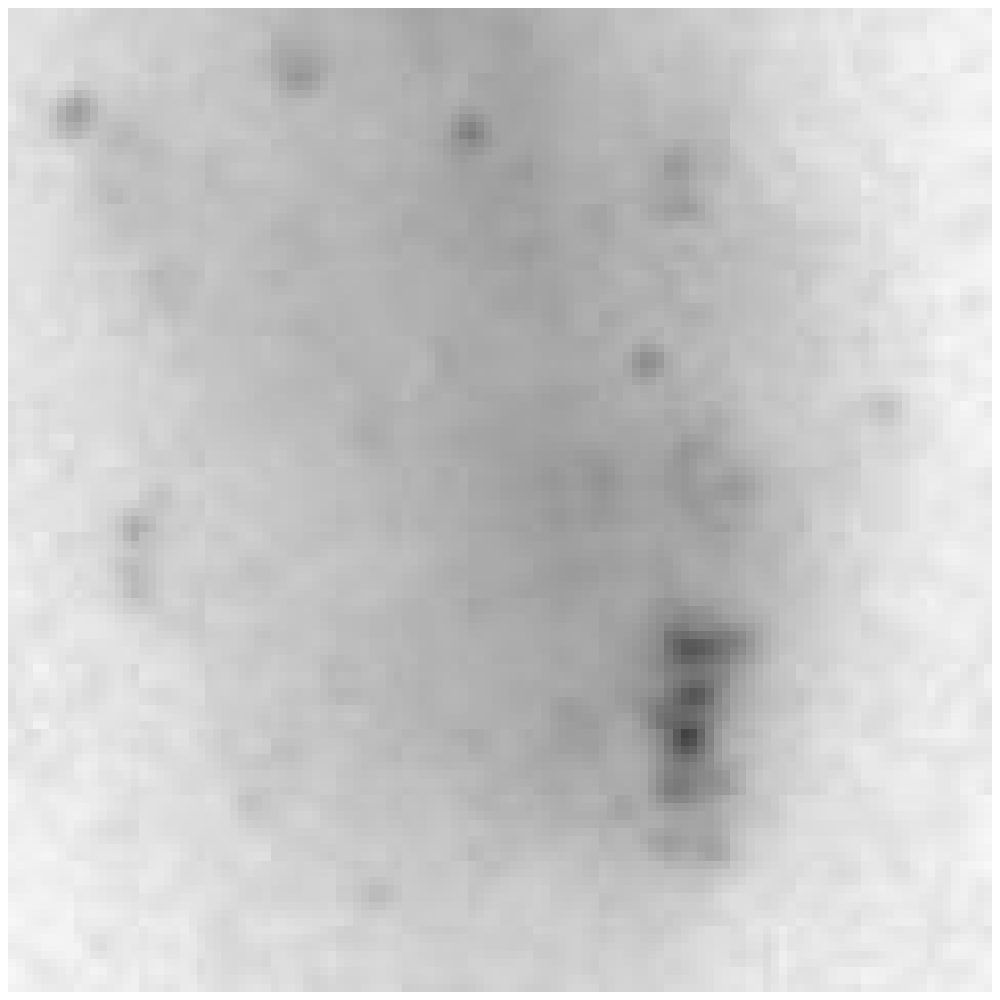}\\
{2MFGC\,10867 = UGC\,8507} \caption{Examples of objects excluded
from 2MFGC. The combined $J + H + K_s$\,\, 2MASS images (on the
left) and DSS2 or SDSS images (on the right).}
%\end{minipage}
\end{figure*}

The 2MFGC\,813 galaxy, given in the first pair of images, shows
almost an ideal case (Fig.~1). The vast majority of visually
selected galaxies are similar to it. The second (2MFGC\,895, $b =
-5\fdg5$) and third (2MFGC\,1119, $b = 0\fdg6$) pairs show
complicated variants of the galaxies' location in the Zone of
Avoidance, where the density of our Galaxy's stars is high. They
were qualified as satisfactory based on the formal selection
principle, however, in the reference for 2MFGC\,895 we noted the
presence of a bright star nearby, and the red galaxy 2MFGC\,1119,
scarcely noticeable in the visible region in the DSS, was included
in the list without any references. The fourth pair (2MFGC\,9495)
is an example of a flat galaxy interacting with another galaxy. In
2MASS, both galaxies are separately detected. Other pairs of
interacting galaxies, such as 2MFGC\,151 (Fig.~2), were excluded
from the list, because when measured photometrically in 2MASS,
both non flat galaxies were combined into one object. This
resulted in the elongated shape of the object, and the position
angle was measured by the line connecting the two galaxies. A
similar asymmetry in the infrared image of an object also appears
with galaxy and (or) star chains, e.g., MFGC\,673, in Fig.~2. The
third pair of images in Fig.~2 (2MFGC\,5959) shows an example when
a galactic bar is seen in 2MASS and the spiral structure is
noticeable only in the visible range. The last pair of images is
an example of the small bright part of the 2MFGC\,10867 galaxy
marked in 2MASS.

The first review of images of all the objects from the 2MFGC
allowed us to detect more than 2000 objects which are asymmetric
or have a nodular structure in the IR range while their shape in
the visible range is more rounded than in the IR range, or they
have a multiplet structure, and some cannot be seen at all. In the
course of further comparison of the 2MASS images ($J, H, K_s$) and
the DSS1, DSS2, and \mbox {SDSS images} of the detected objects
and their vicinities, and also considering their sizes, distances,
location, orientation, and positional angles measured in 2MASS, we
selected 1512 ``false'' objects, which make up 8.4\% of their
total number in the catalog. In the references to the electronic
table (see the footnote on page~\pageref{maintable:Mitronova_n}),
the following abbreviations are used for the remarks why they
cannot be considered flat objects:
\begin{list}{}{
\setlength\leftmargin{4mm} \setlength\topsep{1mm}
\setlength\parsep{0mm} \setlength\itemsep{1mm} }

\item PofG---close pairs of galaxies; %(pair of galaxies);
\item GG---multiple galaxies; %(2 or more galaxies);
\item G\,+\,S---a star which was not excluded from galaxy photometry; %(galaxy and star on line);
\item SS---multiple stars; %(2 or more stars on line);
\item IRbar---only a bar or a bulge of a galaxy is seen in 2MASS; %(in NIR only bar of galaxy);
\item IRcen---only the central part of a galaxy is seen in 2MASS; %(in NIR central part of galaxy);
\item PART---only a certain part of a galaxy is seen in 2MASS; %(part of galaxy);
\item nEon---a galaxy is not flat; %(no edge-on galaxy);
\item asymm---an asymmetric form of a galaxy of unknown nature; %(asymmetrical form of galaxy);
\item Interacting---interacting systems;
\item 2obj, 3obj---multiple objects of small angular sizes;%(interacting systems).
\item MIST---damaged 2MASS images. %(mistake on the 2MASS images);
\end{list}

The objects marked with the \# sign in the catalog references, in
our view, should be excluded from the list. Near the three
``false'' objects with the 2MFGC numbers 3795, 5287, and 14518, we
found several new flat galaxies which have never been mentioned
before in any catalogs including 2MASS XSC, on which the 2MFGC is
based.
%we found flat galaxies not included in 2MASS~XSC, on which the
%2MFGC is based, or in any other catalogs.
Table~1 shows the number of the 2MFGC objects marked in the
catalog with the mentioned abbreviations. About 80~objects were
excluded from the catalog for two or more reasons, e.g., multiple
systems often show tidal interaction, and some pairs are
surrounded by other galaxies which can also be included in 2MASS
photometry. According to our observations, the most ``false''
objects are a result of combined measurements of galaxy pairs,
galaxies with a star (stars), pairs and chains of stars.

%\renewcommand{\baselinestretch}{0.8}
%1
\begin{table}[hbt]
\setcaptionmargin{0mm}
 \onelinecaptionstrue
 \captionstyle{normal}
\caption{The distribution of 1512 objects excluded from the 2MFGC
catalog}
\medskip
\begin{tabular}{l|c|c}
 \hline
 {Reasons}  & Number  & Percentage\\
  & of objects & of the 2MFGC\\
 \hline
PofG, 2PofG, 2G, 2obj    &  644 & 3.58 \\
G\,+\,S                      &  266 & 1.48 \\
GG, GG\,+\,S, GGSS, 3obj, SS &  268 & 1.49 \\
nEon, asymm              &  219 & 1.21 \\
IRbar, IRcore            &   93 & 0.52 \\
Interacting systems      &   81 & 0.45 \\
PART                     &   12 & 0.07 \\
MIST, G\,+\,MIST             &    5 & 0.03 \\
\hline
\end{tabular}
\end{table}
%\renewcommand{\baselinestretch}{1.0}

%\renewcommand{\baselinestretch}{0.85}
%2
\begin{table*}[hbt]
\setcaptionmargin{0mm}
 \onelinecaptionstrue
 \captionstyle{normal}
\caption{The duplicated 2MFGC galaxies with their coordinates and
numbers in LEDA}
\medskip
\begin{tabular}{c|c|c|c|c|c}
\hline
2MFGC & RA (2000.0) Dec &  LEDA  & $dL$ &    2MFGC & RA (2000.0) Dec\\
\hline
(1) & (2)  &  (3)  & (4) & (5) & (6)\\
\hline

1582  & $020404.91$$-$$080726.0$ &1007972 &.2 &  1584  & $020406.97$$-$$080735.5$ \\
3260  & $035821.82$$-$$442758.5$ &  14190 &.1 &  3262  & $035823.35$$-$$442802.7$ \\
3664  & $043026.12$$+$$884615.2$ &  15599 &.1 &  3660  & $043008.43$$+$$884617.8$ \\
4491  & $053058.45$$-$$535237.0$ &  17381 &.3 &  4493  & $053059.38$$-$$535244.9$ \\
5794  & $071651.92$$-$$185234.4$ &   --   &   &  5793  & $071650.96$$-$$185225.2$ \\
5972  & $073200.37$$+$$834256.5$ &2788269 &.1 &  5974  & $073204.07$$+$$834253.2$ \\
6642  & $082502.42$$+$$742558.1$ &  23618 &.1 &  6644  & $082504.09$$+$$742602.1$ \\
7027  & $090014.46$$+$$354352.7$ &  25281 &.4 &  7030  & $090015.77$$+$$354338.8$ \\
7381  & $093104.91$$+$$875310.1$ &   --   &   &  7380  & $093101.92$$+$$875315.8$ \\
8137  & $102941.92$$+$$685041.3$ &2724517 &.0 &  8136  & $102941.45$$+$$685048.3$ \\
8237  & $103621.13$$-$$264609.9$ &  31437 &.2 &  8239  & $103621.46$$-$$264622.9$ \\
8339  & $104238.04$$-$$235608.9$ &  31919 &.1 &  8337  & $104237.46$$-$$235605.7$ \\
8976  & $112825.06$$+$$092427.1$ &  35314 &.2 &  8974  & $112824.04$$+$$092427.8$ \\
9580  & $121034.60$$+$$581814.6$ &  38741 &.2 &  9578  & $121032.87$$+$$581813.8$ \\
9833  & $122844.25$$+$$114540.8$ &  41060 &.5 &  9831  & $122843.62$$+$$114526.1$ \\
11311 & $135948.71$$+$$402248.8$ &  49817 &.2 &  11309 & $135947.71$$+$$402256.5$ \\
11725 & $142725.54$$-$$874610.3$ &  51613 &.1 &  11723 & $142720.46$$-$$874620.1$ \\
11829 & $143451.25$$-$$295654.8$ &  92454 &.0 &  11827 & $143451.02$$-$$295651.1$ \\
12527 & $153122.24$$-$$872616.2$ &  55293 &.1 &  12524 & $153118.31$$-$$872601.8$ \\
14584 & $184721.78$$-$$531214.8$ &   --   &   &  14585 & $184722.12$$-$$531219.8$ \\
15821 & $205411.51$$+$$174657.3$ &  65683 &.2 &  15823 & $205412.19$$+$$174645.6$ \\
15849 & $205629.27$$-$$484348.5$ &  92627 &.0 &  15847 & $205628.93$$-$$484402.4$ \\
17208 & $225308.79$$-$$534247.6$ & 426959 &.0 &  17206 & $225307.92$$-$$534244.7$ \\
\hline
\end{tabular}
\end{table*}

Moreover, in a closer look, 23 galaxies in the 2MASS catalog were
found to be duplicated.
%In the course of the survey, the measurements of 23~galaxies have
%been found to be duplicated 2MASS.
Table~2 shows their 2MFGC numbers (columns~1~and~5), the
corresponding coordinates (columns~2~and~6), galaxy numbers in the
LEDA database~(3), and the accuracy~(4) of identifying (in
fractions of a minute of arc) the first pair of coordinates with
the coordinates of the object from LEDA. In the 2MFGC references,
the second 2MFGC numbers are given corresponding to the considered
galaxy. The sign \# in the references means that we excluded from
the list this one measurement of the two given.

One of the galaxies (2MFGC\,6642$\,\equiv\,$6644) is considered
being nonflat:
% not flat:
the red bar of the galaxy is seen in the IR range, and
the ring around it appears only in the visible range. The number
of real objects in the 2MFGC catalog decreased by 23 and amounted
to 17\,997. Another 1512 objects are excluded as not meeting the
selection criteria of the catalog. As a result, the 2MFGC catalog
consists of 16\,485~flat galaxies which can be used in further
investigations.

Using the images of 2MFGC galaxies, we determined the
morphological types for more than 3900 of them, which are also
available in the references. The vast majority of such galaxies
were not included in formerly known catalogs, i.e., they were
first found in the 2MASS survey. Into that category fall the
objects at low galactic latitudes, which claim our closer
attention, as they were often registered in 2MASS at the detection
threshold.
%Objects at low galactic latitudes fall into that category and
%claim our closer attention, as they were often registered in 2MASS
%at the detection threshold.
In the DSS images they had a low contrast appearance or were not
even detected against the sky background. When stars or galaxies
with comparable angular sizes were projected near a flat galaxy,
we marked them in the references with the signs: +S, +SS or +G,
+GG. The signs~:~and~? in the references show our doubts. About
seventy doubtful objects are left in the catalog, but if deeper
images appear, some of these objects may turn out to be
``false.\!''

\section{DIAGRAMS OF THE DISTRIBUTION OF~``FALSE'' GALAXIES FROM THE~2MFGC~LIST}

To analyze the properties of 1512 excluded galaxies, we plotted
the diagrams of their distribution compared to the total number
($N = 17\,998$) in the 2MFGC.\!\footnote{Both measurements of the
duplicated galaxy with the numbers 2MFGC\,5793 and 2MFGC\,5794 are
included in this list.} In all histograms (Figs.~3--5), we marked
the percentage of the ``false'' galaxies compared to the total
number in a bin. The $K_s$ distribution maximum of these objects
is shifted only by $0\fm5$ to the region of faint objects
(Fig.~3); however, in the same direction (greater than~$13^{\rm
m}$), their fraction considerably increases in comparison with all
the 2MFGC objects, which can be seen from the percentage of
``false'' galaxies.

%3
\begin{figure*}[hbt]
 %\vspace{2mm}
\begin{minipage}{\linewidth}
\setcaptionmargin{5mm} \onelinecaptionsfalse \captionstyle{normal}
\includegraphics[scale=0.5,angle=-90]{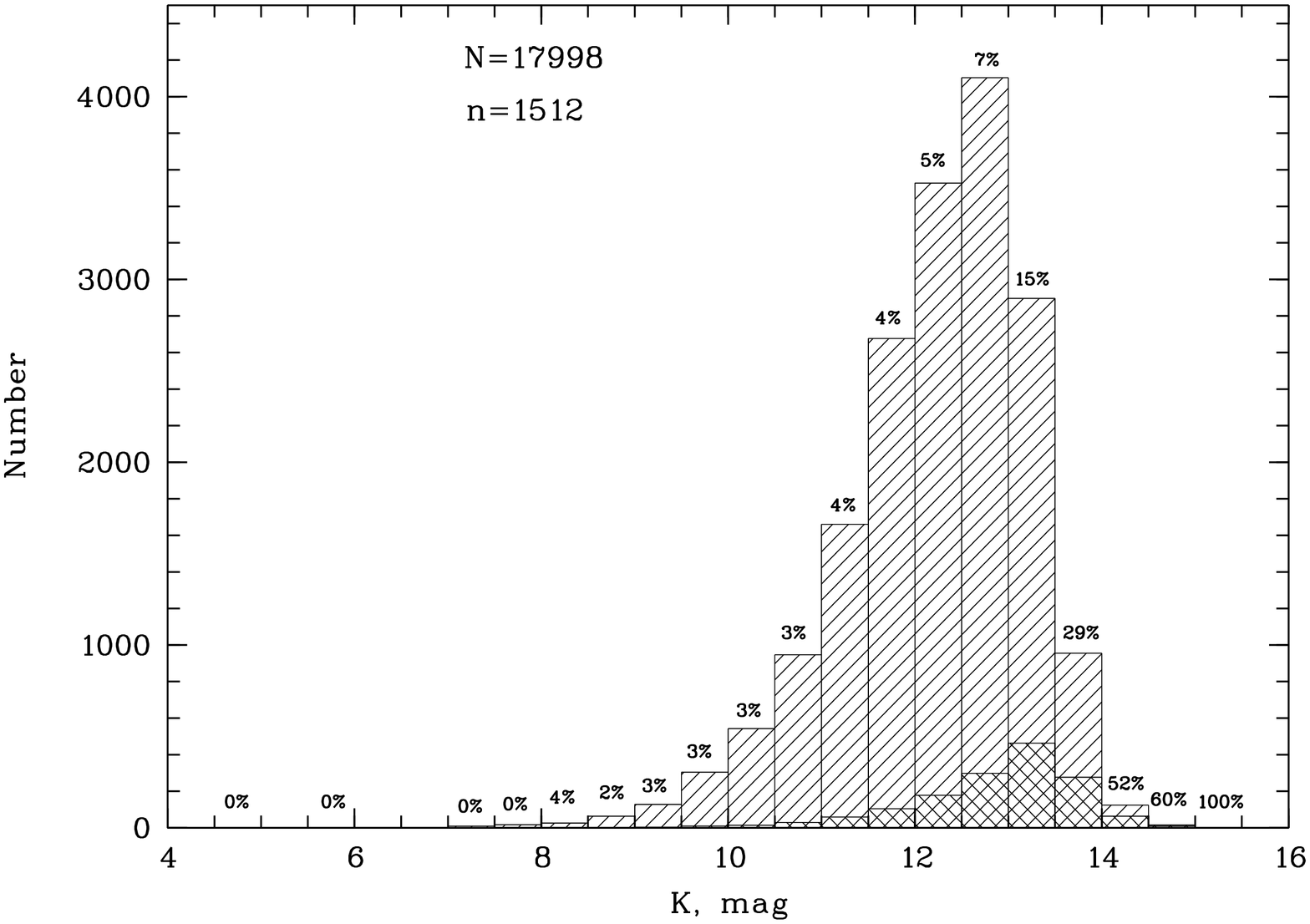}
\caption{The distribution of galaxies by $K_s$. The histogram
shows the percentage of ``false'' galaxies (crosshatched) compared
to the total number (hatched) in each bin, the increment size is
$0\fm5$. The total number of \mbox{2MFGC galaxies} ($N$) and the
number of the ``false'' ones ($n$) are given at the top.}
%\end{figure*}
\end{minipage}
\begin{minipage}{\linewidth}
%4
%\begin{figure*}
\setcaptionmargin{5mm} \onelinecaptionsfalse \captionstyle{normal}
 \vspace{5mm}
\includegraphics[scale=0.5,angle=-90]{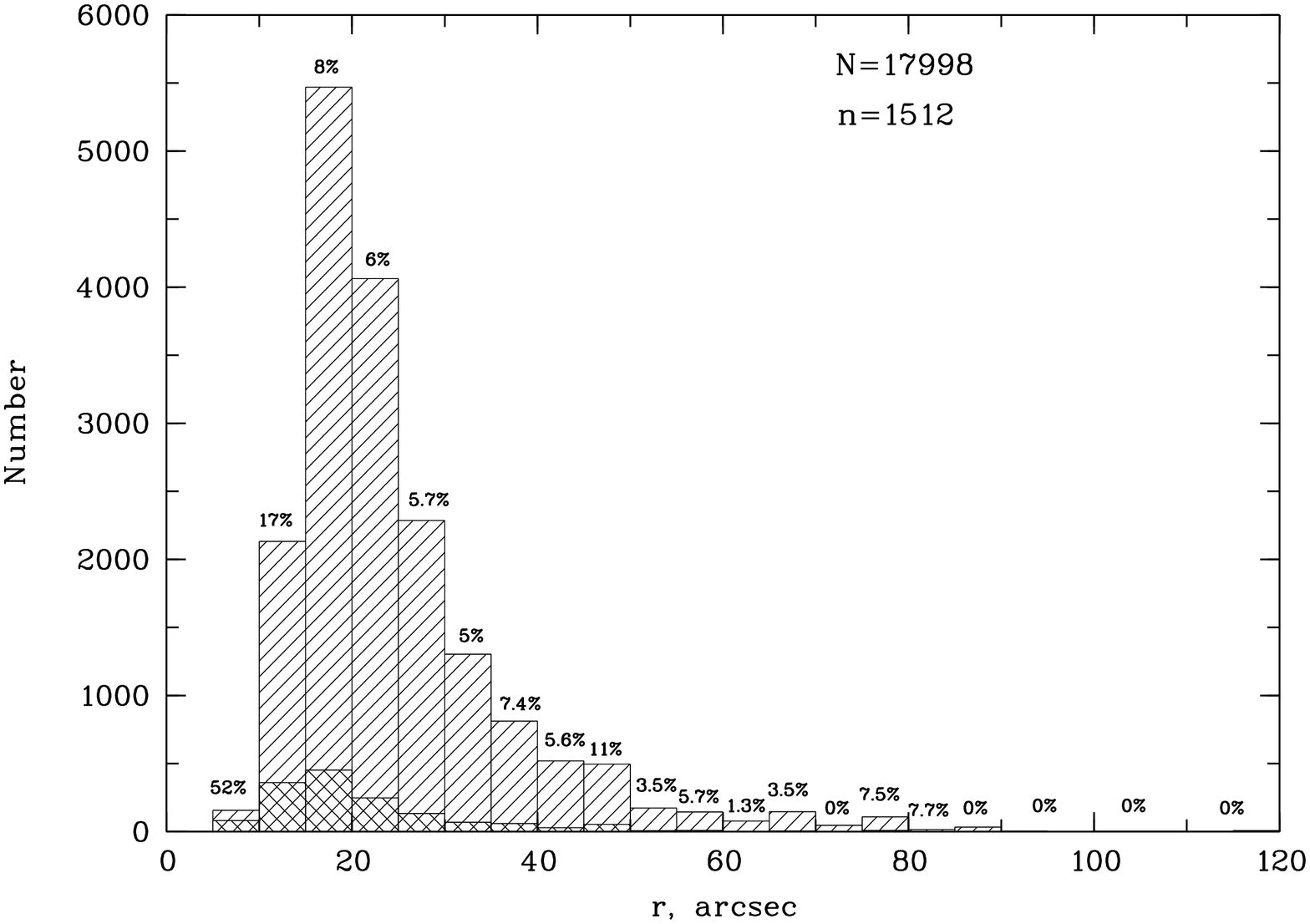}
\caption{The distribution of galaxies by angular diameters. The
percentage of ``false'' galaxies is also shown here, the hatching
is similar to Fig.~3. The total number of \mbox{2MFGC galaxies}
($N$) and the number of ``false'' objects ($n$) are given at the
top.}
% hist_r.ps
\end{minipage}
\end{figure*}

%5
\begin{figure*}
 %\vspace{2mm}
\begin{minipage}{\linewidth}
\setcaptionmargin{5mm} \onelinecaptionsfalse \captionstyle{normal}
\includegraphics[scale=0.5,angle=-90]{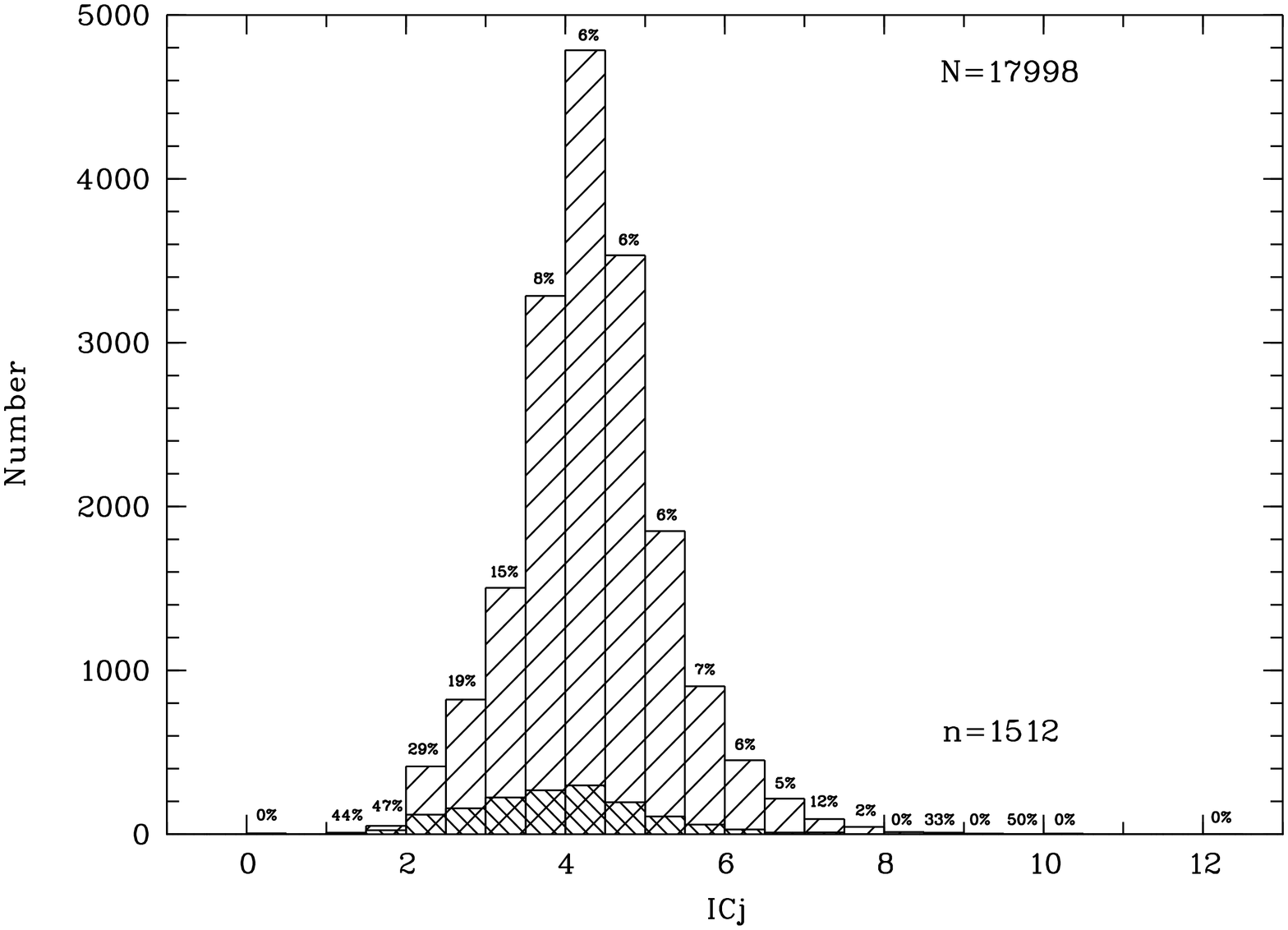}
\caption{The distribution of galaxies by the concentration index.
The percentage of ``false'' galaxies is shown here, the hatching
is similar to Figs.~3 and~4. The total number of \mbox{2MFGC
galaxies} ($N$) and the number of ``false'' objects ($n$) are
given on the right.}
% hist_ICj.ps
%\end{figure*}
\end{minipage}
\begin{minipage}{\linewidth}
%6
%\begin{figure*}
 \vspace{5mm}
\setcaptionmargin{5mm} \onelinecaptionsfalse \captionstyle{normal}
\includegraphics[scale=0.5,angle=-90]{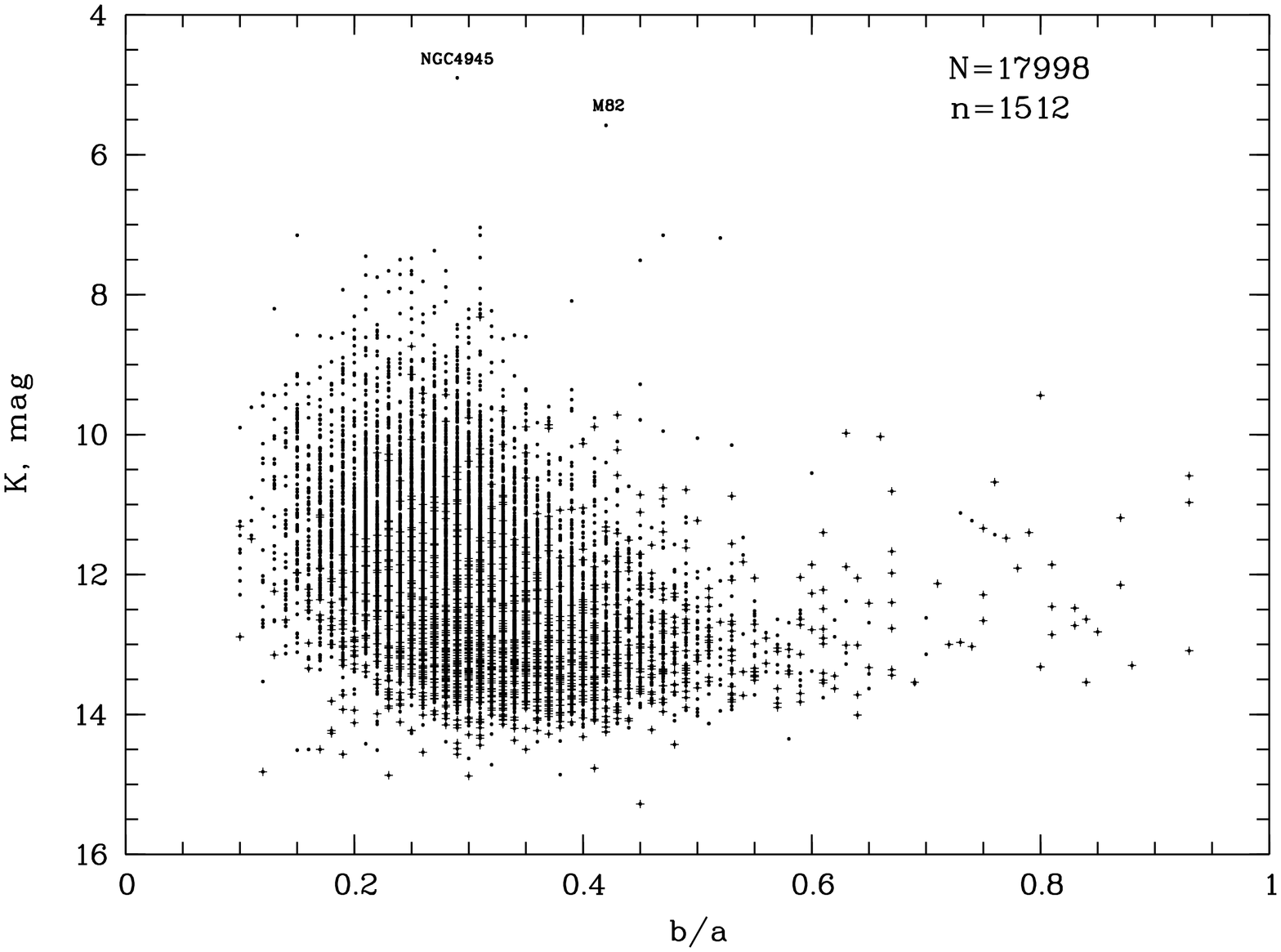}
\caption{The dependence of the $K_s$ magnitude on the axial ratio.
The dots denote all the galaxies, the crosses---the excluded
objects. The total number of \mbox{2MFGC galaxies} ($N$) and the
number of ``false'' objects ($n$) are given at the top.}
% ba_K.ps
\end{minipage}
\end{figure*}

%7
\begin{figure*}
\setcaptionmargin{5mm} \onelinecaptionsfalse \captionstyle{normal}
%\hbox{
%\includegraphics[scale=0.33,angle=-90, bb= 27 47 570 809, clip]{Mitronova_fig7a.eps}\hspace{-2mm}% plot_r_ba1.prg
%\includegraphics[scale=0.33,angle=-90, bb= 27 47 570 809, clip]{Mitronova_fig7b.eps}}
%\vbox{\vspace{5mm}\includegraphics[scale=0.5,angle=0,bb=0 40 792 570,clip]{Mitronova_fig7a.eps}}% plot_r_ba1.prg
%\vbox{\vspace{5mm}\includegraphics[scale=0.5,angle=0,bb=0 40 792
%570,clip]{Mitronova_fig7b.eps}}
\includegraphics[scale=0.5,angle=-90]{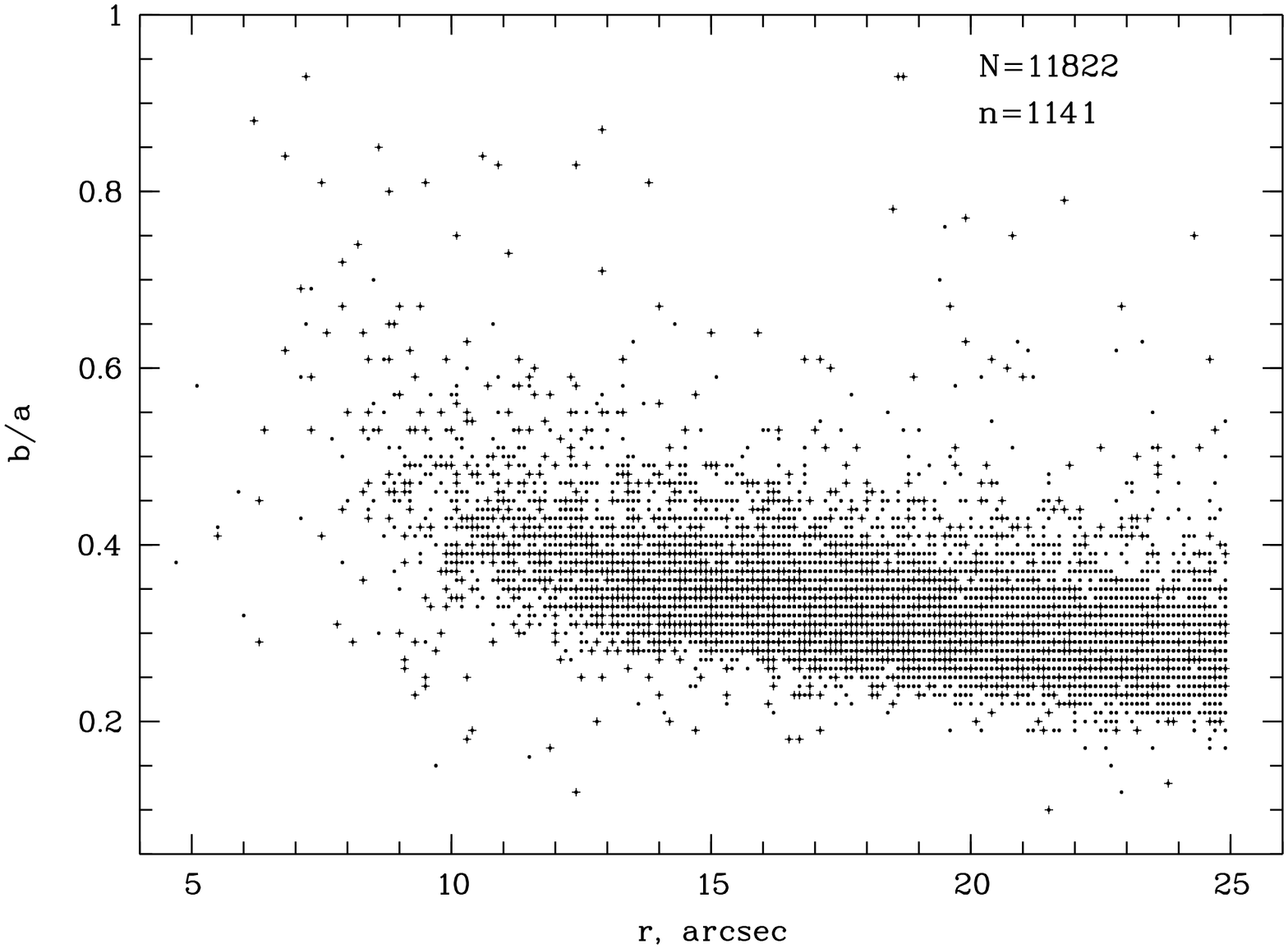}\\
\includegraphics[scale=0.5,angle=-90]{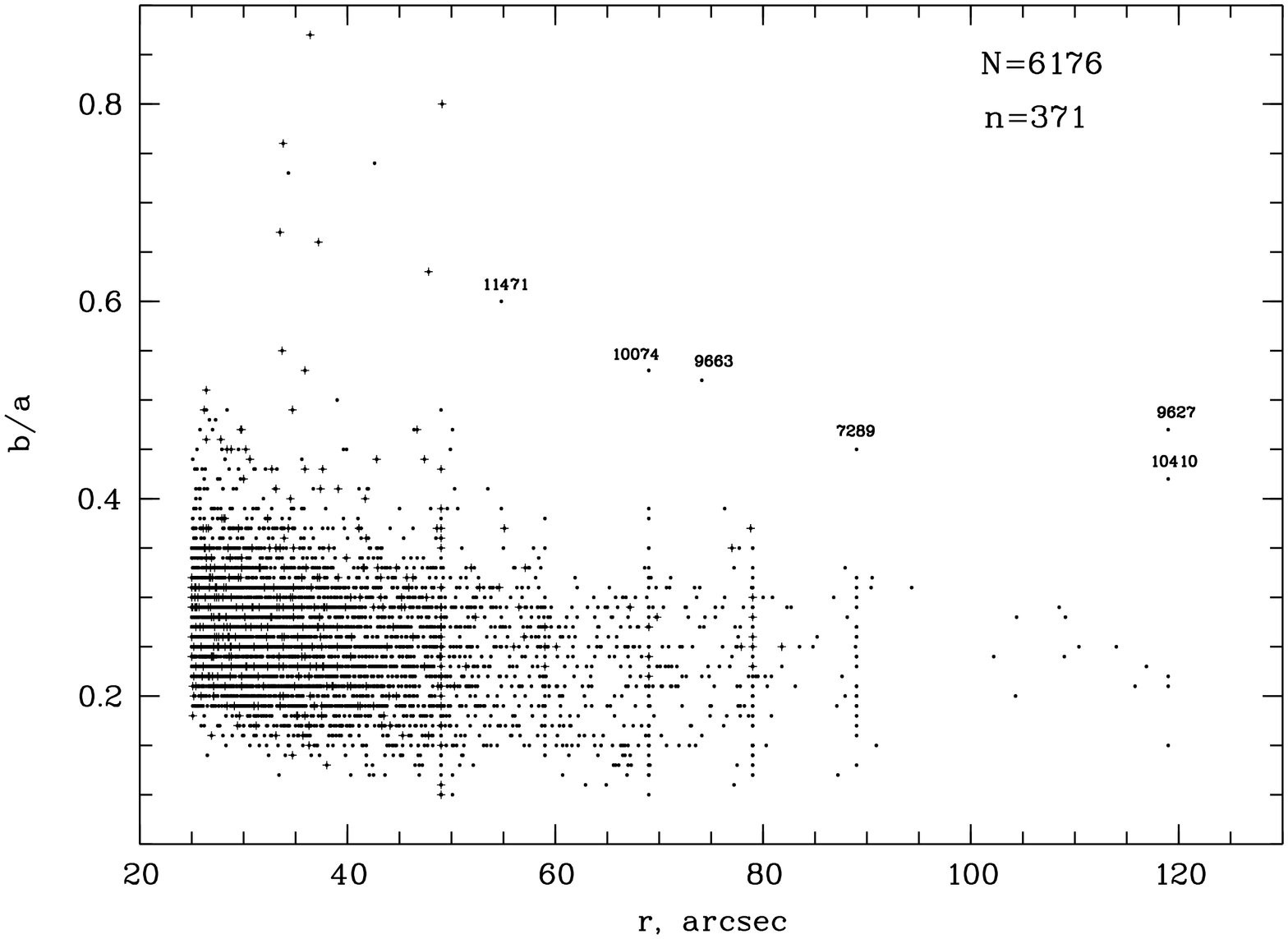}
\caption{The dependence of the axial ratio on the radius $r$ of
galaxies. The upper figure shows all the 2MFGC galaxies with $r <
25\arcsec$, and the lower one---with $r\geq25\arcsec$. The dots
denote all the galaxies, the crosses---the excluded objects. The
number of objects in the intervals is given on top.}
\end{figure*}

The figures show that the percentage of ``false'' objects with
small (\mbox {$r < 10\arcsec$}) angular sizes (Fig.~4) and a low
concentration index $IC_j < 2$ (Fig.~5) turned out to be much
higher, which is not surprising for non deep surveys, to which the
2MASS belongs.

In the two-dimensional distributions of the $K_s$ magnitude as a
function of the axial ratio $b/a$ (Fig.~6) averaged over the
individual $J, H, K_s$ values and the change of this $b/a$ ratio
with the change of Kron elliptical radius $r$ measured by the
twentieth isophote in the $K_s$ filter (Fig.~7), it can be noticed
that the excluded galaxies considerably increased the scatter in
the diagrams. We focused on six galaxies (7289, 9627, 9663, 10074,
10410, 11471) that stay away from the main concentrated cloud on
the lower panel in Fig.~7, for which the actual flatness value is
underestimated in 2MASS~XSC, as judged by the images in 2MASS.

Based on the detailed analysis of the results shown in the
diagrams, we conclude that the objects excluded do not have much
influence on the general distribution of galaxies in the catalog;
although, the scattering of the values decreases, mainly in the
region of the objects with small angular sizes and low surface
brightnesses, where the 2MASS detection threshold has its impact.

\begin{acknowledgements}

We thank Prof.~I.~D.~Karachentsev and D.~I.~Maka\-rov for their
helpful advice and discussions. We are grateful for the
opportunity to use the data from the Two Micron All-Sky Survey
(2MASS) and the Sloan Digital Sky Survey (SDSS-III~DR9). We thank
NASA/IPAC Extragalactic Database (NED) and ESO Online Digitized
Sky Survey for the extensive and continuous operation of the
systems, which were of invaluable help in quick browsing for the
data on the selected objects. The investigation was conducted with
the financial support of the grant from the Russian Scientific
Foundation (project No.\,\mbox{14-12-00965}).

\end{acknowledgements}
%\onecolumn
%\section{REFERENSES}

%\twocolumn
{}
\end{document}